\documentclass[aps,twocolumn,nofootinbib]{revtex4}

\usepackage{amsmath,amssymb}
\usepackage{graphicx}
\newcommand{\amp}{&\!\!}

\newcommand{\Mpl}{M_{\mbox{\tiny{Pl}}}}
\newcommand{\beq}{\begin{equation}}
\newcommand{\eeq}{\end{equation}}
\newcommand{\bea}{\begin{eqnarray}}
\newcommand{\eea}{\end{eqnarray}}
\newcommand{\nn}{\nonumber}

\begin{document}

\title{A Correlation Between the Higgs Mass and Dark Matter}

\author{ Mark~P.~Hertzberg$^{1,2}$}
\affiliation{$^1$Center for Theoretical Physics and Dept.~of Physics,\\ Massachusetts Institute of Technology, Cambridge, MA 02139, USA
\\ $^2$Institute of Cosmology and Dept.~of Physics and Astronomy,\\ Tufts University,\\ Medford, MA 02155, USA}


\begin{abstract}
Depending on the value of the Higgs mass, the Standard Model acquires an unstable region at large Higgs field values due to RG running of couplings, which we evaluate at 2-loop order. For currently favored values of the Higgs mass, this renders the electroweak vacuum only meta-stable with a long lifetime. We argue on statistical grounds that the Higgs field would be highly unlikely to begin in the small field meta-stable region in the early universe, and thus some new physics should enter in the energy range of order, or lower than, the instability scale to remove the large field unstable region. We assume that Peccei-Quinn (PQ) dynamics enters to solve the strong CP problem and, for a PQ-scale in this energy range, may also remove the unstable region. We allow the PQ-scale to scan and argue, again on statistical grounds, that its value in our universe should be of order the instability scale, rather than (significantly) lower. Since the Higgs mass determines the instability scale, which is argued to set the PQ-scale, and since the PQ-scale determines the axion properties, including its dark matter abundance, we are led to a correlation between the Higgs mass and the abundance of dark matter. We find the correlation to be in good agreement with current data.
\end{abstract}

\vspace*{-14.2cm} {\hfill MIT-CTP 4403\,\,\,\,}


\maketitle

\newpage

\section{Introduction}
\let\thefootnote\relax\footnotetext{Email: {\tt mphertz@mit.edu,  mark.hertzberg@tufts.edu}}

Recent LHC results are consistent with the predictions of the Standard Model, including the presence of a new boson that appears to be the Higgs particle  with a mass $m_H\sim 125 - 126$\,GeV \cite{ATLAS,CMS} (more recent measurements are summarized in \cite{Aad:2015zhl}). With the Higgs at this mass, the Standard Model is well behaved up to very high energies if we evolve its couplings under the renormalization group (RG) equations. By no means does this imply that the Standard Model will be valid to these very high energies, and in fact there are good phenomenological reasons, such as dark matter, strong CP problem, baryogenesis, inflation,  hierarchy problem, etc, to think it will be replaced by new physics at much lower energies, say $\mathcal{O}$(TeV). But it is logically possible, albeit unlikely, that the Standard Model, or at least the Higgs sector, will persist to these very high energies and the explanation of these phenomena will be connected to physics at these high, or even higher, energy scales. 

So at what energy scale must the Standard Model breakdown? Obviously new physics must enter by the Planck scale $\Mpl$ where quantum gravity requires the introduction of new degrees of freedom. However, the RG running of the Higgs self-coupling $\lambda$ can dictate the need for new physics at lower energies, depending on the starting value of $\lambda$.
The Higgs mass is related to the self-coupling by $m_H=\sqrt{2\lambda}\,v_{EW}$, where the Higgs VEV is $v_{EW}\approx 246$\,GeV.
For moderate to high values of the Higgs mass, the initial value of $\lambda$, defined at energies of order the electroweak scale, is large enough that it never passes through zero upon RG evolution. On the other hand, for small enough values of the Higgs mass, the self-coupling $\lambda$ passes through zero at a sub-Planckian energy, which we denote $E^*$, primarily due to the negative contribution to the beta function from the top quark, acquiring an unstable region at large field values \cite{Sher:1993mf,Casas:1994qy}. 
The latter occurs for a light Higgs as has been observed. One finds that this renders the electroweak vacuum only meta-stable with a long lifetime. However, we will argue in this paper that it is highly unlikely for the Higgs field in the early universe to begin in the meta-stable region as that would require relatively small field values as initial conditions. Instead it would be much more likely to begin at larger field values, placing it in the unstable region. Hence, the energy scale $E^*$ sets the maximum energy scale for new physics beyond the Standard Model to enter.

There are many possible choices for the new physics. One appealing possibility is supersymmetry, which alters the running of the Higgs self-coupling due to the presence of many new degrees of freedom, likely entering at much lower energies, conceivably $\mathcal{O}$(TeV), or so.
In addition to possibly stabilizing the Higgs potential, supersymmetry can also alleviate the hierarchy problem, improve unification of gauge couplings, and fit beautifully into fundamental physics such as string theory. So it is quite appealing from several perspectives. It is conceivable, however, that even if supersymmetry exists in nature, it is spontaneously broken at very high energies, and in such a scenario we would be forced to consider other possible means to stabilize the Higgs potential.

\begin{figure}[t]
\center{\includegraphics[width=\columnwidth]{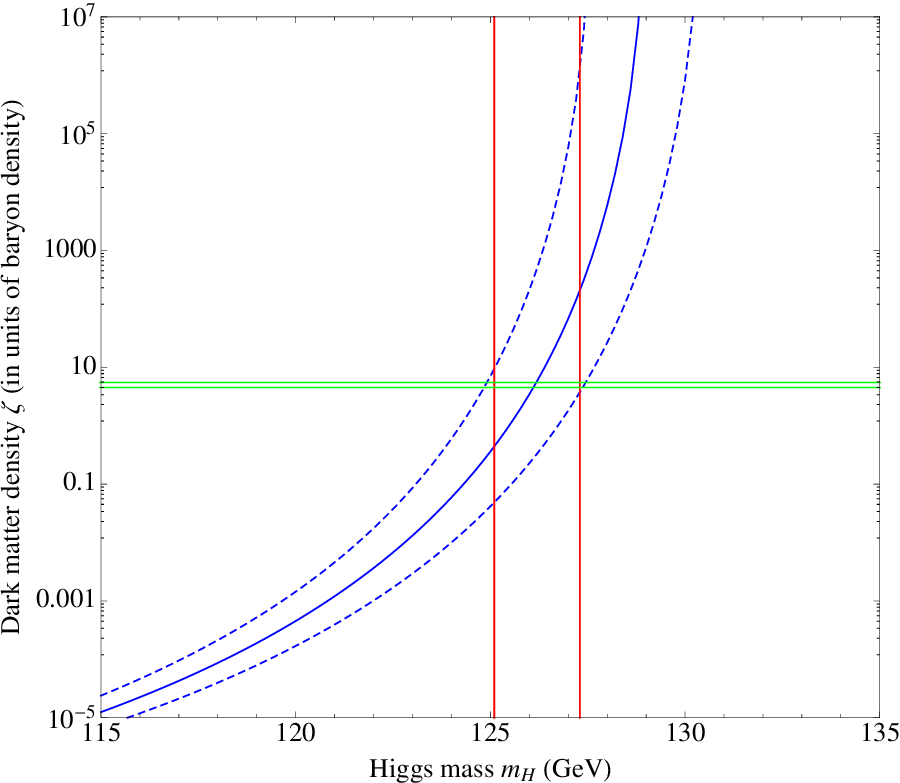}}
\caption{Dark matter density $\zeta$ (in units of baryon density) as a function of the Higgs mass $m_H$ is given by the blue curves. The  solid-blue curve is for the central value of the top mass $m_t=173.1$\,GeV. The dashed-blue curves are for $m_t=173.1\pm 0.7$\,GeV, with the upper value on the right and the lower value on the left.
The red vertical lines indicate the measured Higgs mass range $m_H=125.7\pm 0.6$\,GeV from combining ATLAS and CMS data \cite{ATLAS,CMS}. The green horizontal lines indicate the measured dark matter density to baryon density $\zeta=\Omega_{DM}/\Omega_B$ range, where $\Omega_{DM}=0.229\pm0.015$ and $\Omega_B=0.0458\pm0.0016$, from WMAP7 data  \cite{WMAP7}.}
\label{HiggsAxionPlot}\end{figure}

One intriguing possibility that we examine in this paper is to utilize dynamics associated with the solution of the strong CP problem; the problem that the CP violating term $\sim \theta\,\epsilon^{\mu\nu\alpha\beta}F_{\mu\nu}^aF_{\alpha\beta}^a$ in the QCD Lagrangian is experimentally constrained to have coefficient $|\theta|\lesssim 10^{-10}$, which is highly unnatural.
The leading solution involves new Peccei-Quinn (PQ) dynamics \cite{PQ1977}, involving a new complex scalar field $\phi$ and a new global $U(1)$ symmetry that is spontaneously broken at some energy scale $F_{PQ}$. 
This leads to a new light scalar field known as the axion \cite{WeinbergAxion,WilczekAxion}.
Since it is bosonic, the $\phi$ field adds a positive contribution to the effective $\lambda$ for the Higgs, potentially removing the unstable region, depending on the scale $F_{PQ}$. This elegant mechanism to remove the unstable region was included in the very interesting Ref.~\cite{EliasMiro:2012ay}, where this and other mechanisms were discussed, and was a source of motivation for the present work (also related is \cite{Lebedev:2012zw,Salvio:2015cja,Ballesteros:2016euj}).

In the present paper, we would like to take this elegant mechanism for vacuum stability and push it forward in several respects. Firstly, as already mentioned, we will argue on statistical grounds why the meta-stable vacuum requires stabilization. Secondly, we will allow the PQ-scale to scan and argue, again on statistical grounds, why it should be of order the instability scale $E^*$, rather than orders of magnitude lower. Finally, we will furnish a correlation between the Higgs mass and the axion dark matter abundance, and use the latest LHC \cite{ATLAS,CMS} and cosmological data \cite{WMAP7} to examine the validity of this proposal. The outcome of this series of arguments and computation is presented in Fig.~\ref{HiggsAxionPlot}, which is the primary result of this work.

The outline of our paper is as follows: 
In Section \ref{RG} we examine the running of the Standard Model couplings at 2-loop order. 
In Section \ref{Meta} we examine the meta-stability of the Standard Model vacuum and argue that it is statistically unfavorable for the Higgs to begin in this region.
In Section \ref{PQDynamics} we include Peccei-Quinn dynamics to remove the Higgs instability and argue why $F_{PQ}\sim E^*$.
In Section \ref{AxionDM} we relate the PQ-scale to the axion dark matter abundance, which furnishes a correlation between the Higgs mass and the abundance of dark matter.
Finally, in Section \ref{Discussion} we compare the correlation to data and discuss our results.

\section{Standard Model RG Evolution}\label{RG}

We begin with a reminder of the structure of the Higgs sector of the Standard Model. 
The Higgs field is a complex doublet $H$ with Lagrangian 
\beq
\mathcal{L}=D H^\dagger D H+\mu^2 H^\dagger H-\lambda(H^\dagger H)^2.
\eeq
In the unitary gauge we expand around the VEV as $H=(0,v_{EW}+h)/\sqrt{2}$, where in our convention $v_{EW}\approx 246$\,GeV. The associated Higgs mass is $m_H=\sqrt{2\lambda}\,v_{EW}$ in terms of the starting value of $\lambda$, normally defined around the $Z$ boson mass.
At higher energies, the self-coupling $\lambda$ undergoes RG evolution due to vacuum fluctuations from self interaction, fermion interactions, and gauge interactions. Defining $d\lambda/dt=\beta_\lambda$ with $t=\ln E/\mu$
the associated 1-loop beta function (suppressing external leg corrections for now) is
\beq
\beta_\lambda = {1\over(4\pi)^2}  \left[24 \lambda ^2-6 y_t^4+\frac{3}{8} \left(2 g^4+\left(g^2+g'^2\right)^2\right)\right],
\eeq
where the only fermion Yukawa coupling we track is that of the top quark $y_t$ since it is by far the largest.
 For sufficiently large Higgs mass, the positive self interaction term $\sim +\lambda^2$ is large enough to keep the beta function positive, or only slightly negative, to avoid $\lambda$ running negative at sub-Planckian energies.
For sufficiently small Higgs mass, the negative top quark contribution $\sim -y_t^4$ can dominate and cause the beta function to go negative, in turn causing $\lambda$ to pass through zero at a sub-Planckian energy, which we denote $E^*$. The top quark Yukawa coupling itself runs toward small values at high energies with 1-loop beta function
\beq
\beta_{y_t} = {y_t\over(4\pi)^2} \left[-\frac{9}{4} g^2-\frac{17}{12}g'^2-8 g_s^2+\frac{9}{2}  y_t^2\right],
\eeq
which is quite sensitive to the value of the strong coupling $g_s$.
To compute the evolution of couplings and the quantity $E^*=E^*(m_H,y_t,\ldots)$ accurately, we do the following: (i) Starting with couplings defined at the $Z$ mass, we perform proper pole matching and running up to the top mass, (ii) we include external leg corrections (and the associated wavefunction renormalization), (iii) we simultaneously solve the 5 beta function differential equations for the 5 important couplings $\lambda, y_t, g',g,g_s$, and (iv) we include the full 2-loop beta functions for the Standard Model; these are presented in the Appendix (see Refs.~\cite{Ford,Luo:2002ey} for more information). 
In our numerics, we use particular values of the couplings $g',g,g_s$, derived from the best fit values 
\beq
\alpha(m_Z)={1\over127.9},\,\sin^2\theta_W=0.2311,\,\alpha_s(m_Z)=0.1184.
\eeq 
In our final analysis, we will allow for three different values of $m_t=\sqrt{2}\,y_t\,v_{EW}$, namely the central value and 1-sigma variation $m_t=173.1\pm 0.7$\,GeV, and we will explore a range of $m_H=\sqrt{2\lambda}\,v_{EW}$, with $v_{EW}=246.22$\,GeV.

\begin{figure}[t]
\center{\includegraphics[width=\columnwidth]{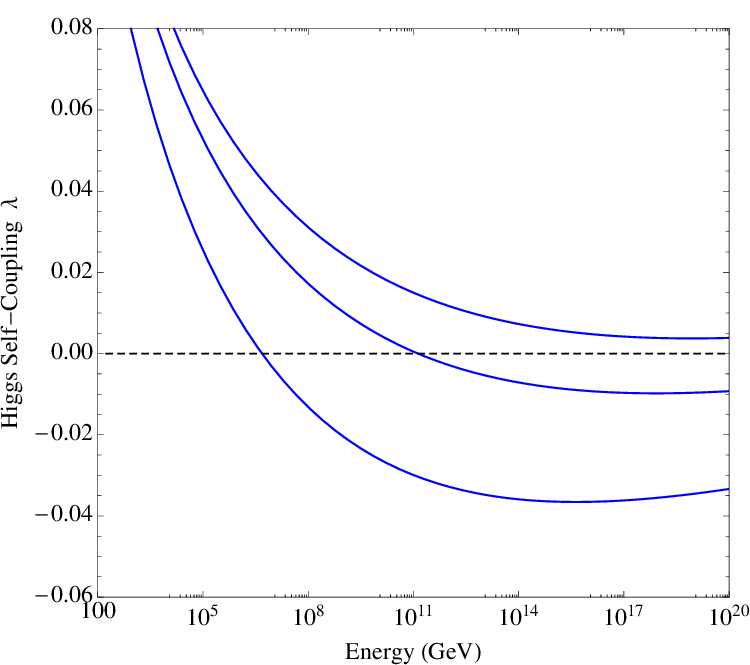}}  
\caption{Higgs self-coupling $\lambda$ as a function of energy, for different values of the Higgs mass from 2-loop RG evolution. 
Lower curve is for $m_H=116$\,GeV, middle curve is for $m_H=126$\,GeV, and upper curve is for $m_H=130$\,GeV.
All other Standard Model couplings have been fixed in this plot, including the top mass at $m_t=173.1$\,GeV.}
\label{LambdaPlot}\end{figure}

Performing the RG evolution leads to the energy dependent renormalized coupling $\lambda(E)$. 
A plot of $\lambda(E)$ is given in Fig.~\ref{LambdaPlot} for three Higgs mass values, namely $m_H=116$\,GeV (lower curve), $m_H=126$\,GeV (middle curve), and $m_H=130$\,GeV (upper curve), with the top mass fixed to the central value $m_t=173.1$\,GeV.
This shows clearly that for the lighter Higgs masses that the coupling $\lambda$ passes through zero at a sub-Planckian energy scale $E^*$
and then remains negative. Furthermore, since the coupling only runs logarithmically slowly with energy, the value of $E^*$ can change by orders of magnitude if the starting value of the couplings changes by relatively small amounts. The domain $E>E^*$ involves a type of ``attractive force" with negative potential energy density, as we now examine in more detail.

\section{Meta-Stability and Probability}\label{Meta}

If we think of the field value $h$ as being the typical energy pushed into a scattering process at energy $E$, then we can translate the RG evolution of the couplings into an effective potential.
Using $\lambda(E)$ and replacing $E\to h$, we obtain the (RG improved) effective potential at high energies ($h\gg v_{EW}$) (see Ref.~\cite{Buttazzo:2013uya} for a precise analysis)
\beq
V_{\tiny{\mbox{eff}}}(h)={1\over4}\lambda(t)\,G(t)^4\,h^4, 
\eeq
where the wavefunction renormalization factor $G$ is given in terms of the anomalous dimension $\gamma$ by $G(t)=\exp(-\int_0^t \gamma(t')dt')$, and we replace $t\to \ln h/\mu$. 
Hence for a Higgs mass in the range observed by the LHC, the effective potential $V_{\tiny{\mbox{eff}}}$ goes negative at a field value $h=E^*$ that is several orders of magnitude below the Planck scale, as can be deduced from the behavior of $\lambda(E)$ with 
$m_H=126$\,GeV in Fig.~\ref{LambdaPlot}.

\begin{figure}[t]
\center{\includegraphics[width=\columnwidth]{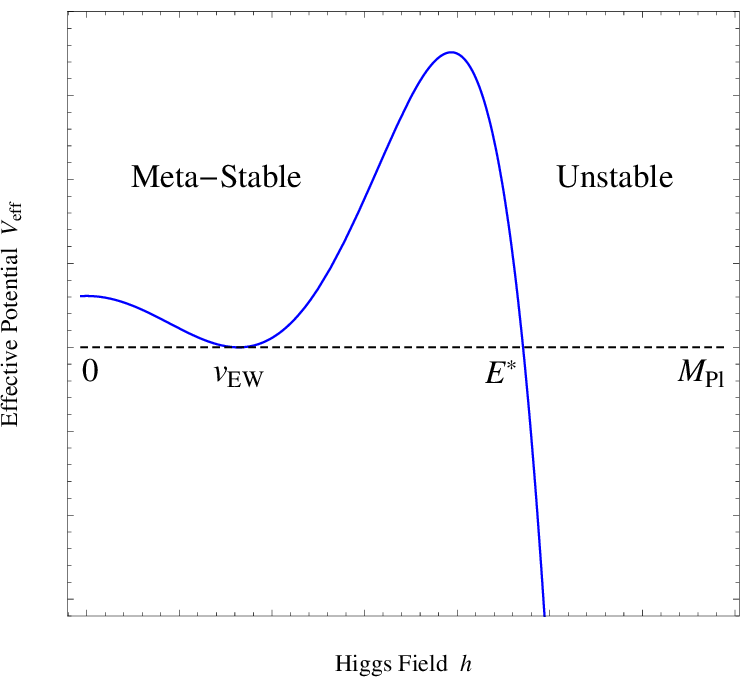}}  
\caption{Schematic of the effective potential $V_{\tiny{\mbox{eff}}}$ as a function of the Higgs field $h$.  This is {\em not} drawn to scale; for a Higgs mass in the range indicated by LHC data, the heirarchy is $v_{EW}\ll E^*\ll \Mpl$, where each of these 3 energy scales is separated by several orders of magnitude.}
\label{VPlot}\end{figure}

We could plot $V_{\tiny{\mbox{eff}}}(h)$ directly, however the factor of $h^4$ makes it vary by many orders of magnitude as we explore a large field range. Instead a schematic of the resulting potential will be more illuminating for the present discussion in order to highlight the important features, as given in Fig.~\ref{VPlot}.
The plot is not drawn to scale; the 3 energy scales satisfy the hierarchy $v_{EW}\ll E^*\ll \Mpl$ for a Higgs mass as indicated by LHC data $m_H\sim 125-126$\,GeV. Note that the local maximum in the potential occurs at a field value that is necessarily very close to $E^*$ (only slightly smaller) and so we shall discuss these 2 field values interchangeably.

In this situation, the electroweak vacuum is only meta-stable. Its quantum mechanical tunneling rate can be estimated 
by Euclideanizing the action and computing the associated bounce action $S_0$. This leads to the following probability of decaying in time $T_U$ through a bubble of size $R$ \cite{Isidori:2001bm}
\beq
p \sim (T_U/R)^4 e^{-S_0}.
\eeq
The computation of the rate is rather involved, and we shall not pursue the details here. Suffice to say that for the central values of Higgs mass and top mass from LHC data,  it is found that the lifetime of the electroweak vacuum is longer than the present age of the universe \cite{Degrassi:2012ry,Masina:2012tz}. 

It is conceivable that it is an acceptable situation for the electroweak vacuum to be meta-stable. However, here we would like to present an argument that such a situation is statistically disfavorable. We imagine that in the very early universe, the Higgs field was randomly distributed in space. For instance, during cosmological inflation the Higgs field could have been frozen at some value as the universe rapidly expands (if high scale inflation) until after inflation when the field will oscillate and its initial value could plausibly have been random and uniformly distributed. If this is the case, then what is the probability that the Higgs field began in the meta-stable region $h\lesssim E^*$, rather than the unstable region $h\gtrsim E^*$? The answer depends on the allowed domain the Higgs can explore. Here we estimate the allowed domain to be Planckian, i.e., $0<h<\Mpl$, but our argument only depends on the upper value being much larger than $E^*$.
Naively, this would lead to a probability $\sim E^*/\Mpl$, however we should recall that the Higgs is a complex doublet, composed of 4 real scalars, and each one would need to satisfy $h\lesssim E^*$ in the early universe to be in the meta-stable region. Hence, we estimate the probability as
\beq
\mbox{Prob}\,(\mbox{Higgs begins in meta-stable region})\sim\left(E^*\over\Mpl\right)^{\!\!4}.
\label{Prob1}\eeq
For example, if we describe the physics in Coulomb gauge, then we have both the modulus of the Higgs field $h$, plus angular modes $\theta_a$, with $a=1,2,3$. From this point of view, it seems most reasonable to take the probability density weighted by an appropriate Jacobian factor associated with transforming from cartesian field co-ordinates to such radial plus angular co-ordinates. This Jacobian scales as $\sim h^3$, and so again will lead to the probability growing like the fourth power of the energy. Another way to put it is to say that there is much more field space available at large Higgs fieldÕs values than for small values. This seems reasonable, especially if one imagines initial conditions laid down by inflation. The number of states in the Hilbert space whose typical Higgs value is large, is much greater than the number of states in the Hilbert space whose typical Higgs value is small. One might reach a different perspective in, say, unitary gauge where the angular modes appear as the longitudinal modes of the $W^{\pm}$ and $Z$ bosons. However, the unitary gauge is not a useful way of describing physics above the electroweak scale. So we consider the above point of view with multiple scalars to be more physically reasonable.

So for instance, for $m_H\approx 125.5$\,GeV and $m_t=173.1$\,GeV, we have $E^*\sim 10^{11}$\,GeV, leading to a probability $\sim (10^{11}\,\mbox{GeV}/10^{19}\,\mbox{GeV})^4=10^{-32}$,
which indicates that the chance of randomly landing in the meta-stable region in the early universe is exceedingly unlikely.
Instead it is far more likely to land in the unstable region indicated in Fig.~\ref{VPlot}. 
Here the effective potential is negative leading to a catastrophic  runaway instability, perhaps to a new VEV that is close to Planckian. This would in turn lead to a plethora of problems for the formation of complex structures, etc, so we can safely assume such a regime is uninhabitable and irrelevant.
This leads us to examine a scenario in which new physics enters and removes this problem.

\section{Peccei-Quinn Dynamics and Distribution}\label{PQDynamics}

One of the phenomenological reasons for new physics beyond the Standard Model is the fine tuning of the CP violating term in the QCD Lagrangian.
The following dimension 4 operator is gauge invariant and Lorentz invariant and should be included in the QCD Lagrangian with a dimensionless coefficient $\theta$
\beq
\Delta\mathcal{L}={\theta\over 32\pi^2}\epsilon^{\mu\nu\alpha\beta}F_{\mu\nu}^aF_{\alpha\beta}^a \,.
\eeq
From bounds on the electric dipole moment of the neutron, this term is experimentally constrained to satisfy $|\theta|\lesssim 10^{-10}$, which requires extreme fine tuning. There appears to no statistical explanation of this fine tuning if it were purely random, since a small but moderate value of $\theta$ would have very little consequences for the formation of complex structures. Instead this requires a dynamical explanation, which we take to be due to a new global symmetry, known as a Peccei-Quinn (PQ) symmetry \cite{PQ1977}, involving a new heavy complex scalar field $\phi$. This field undergoes spontaneous symmetry breaking at a scale $F_{PQ}$, and in the resulting effective field theory, the quantity $\theta$ is essentially promoted to the angular degree of freedom in $\phi=\rho \, e^{i\theta}/\sqrt{2}$; a light scalar field known as the axion \cite{WeinbergAxion,WilczekAxion}.  The zero temperature Lagrangian for $\phi$ includes a symmetry breaking potential for $\rho$ and a QCD instanton generated sinusoidal potential for $\theta$
\bea
\mathcal{L}\amp=\amp{1\over2}\rho^2(\partial_\mu\theta)^2+{1\over2}(\partial_\mu\rho)^2\nonumber\\
\amp\amp-2\Lambda^2\sin^2(\theta/2)-{\lambda_\rho\over4}(\rho^2-F_{PQ}^2)^2 .
\eea
If we expand around the $\phi$-field's VEV at low energies, we see that the angular component of $\phi$, the axion, is very light with a mass $m_a= \Lambda^2/F_{PQ}$ (where $\Lambda$ is of order the QCD-scale) and will be dynamically driven to zero, solving the strong CP problem.
Since we will require $F_{PQ}$ to be a very high energy scale (compared to say the electroweak scale), the radial mode of $\phi$ is very heavy, with mass $\sim F_{PQ}$.
Hence at low energies, the radial mode is essentially irrelevant; it can be integrated out and, apart from a possible renormalization of the Standard Model couplings, can be ignored. However at energies approaching $F_{PQ}$, the radial mode cannot be ignored.
The field $\phi$ couples to various Standard Model particles in any realization of the PQ symmetry, including the Higgs field through the interactions of the type $\sim \lambda_{h\rho}\,H^\dagger H \phi^*\phi$ (where $\lambda_{h\rho}$ is a dimensionless coupling).
This causes an alteration in the effective coupling $\lambda$ at an energy scale of order $E\sim m_\rho$ where the new field becomes dynamical.

Since $\phi$ is bosonic, it generally leads to a positive increase in $\lambda$, either through tree-level corrections or through loop corrections as follows: as long as $\lambda_{h\rho}$ is not very small, then it makes a significant and rapid change in the $\beta$-function for $\lambda$ of the form $\Delta\beta_\lambda\sim\lambda_{h\rho}^2$. This is because in the cases of interest, $\lambda$ is otherwise very small in the vicinity of this effect turning on, as seen in Fig. 2. So even a small positive change in its $\beta$ function can cause a rapid change and stabilization of the effective potential
 leading to a threshold boost in $\lambda$ at the scale at which this new degree of freedom becomes active; a point that was included nicely in Ref.~\cite{EliasMiro:2012ay}. This conclusion can only be avoided by an atypically tiny coupling between the Higgs and the PQ-field.

In the most common case then, this leads to a reduction or removal of the unstable region depending on the scale $F_{PQ}$ relative to the instability scale $E^*$, assuming $\mathcal{O}(1)$ couplings between the Peccei-Quinn dynamics and the Higgs sector.
The more precise statement is that the mass of the radial field is $m_\rho=\sqrt{\lambda_\rho}\,F_{PQ}$. This really sets the scale at which a correction to the $\beta$ function becomes active.

We obviously require $F_{PQ}$ to be in the range $m_\rho=\sqrt{\lambda_\rho}\,F_{PQ}\lesssim E^*$ in order for the new physics to prevent the effective potential $V_{\tiny{\mbox{eff}}}(h)$ from having a large negative regime (note that a small negative dip is statistically allowable for $h$, but not a large field dip). But since $E^*$ is very large, this leaves several orders of magnitude uncertainty in the value of $F_{PQ}$. In other words, it would be sufficient for $F_{PQ}\ll E^*$ in order to remove the unstable region. However, here we would like to present a statistical argument that 
\beq
\sqrt{\lambda_\rho}\,F_{PQ}\sim E^*
\eeq 
is much more likely. We shall take $\lambda_\rho=\mathcal{O}(1)$ in the following discussion to illustrate the idea, though it is simple to generalize the argument.
There are indications that the PQ-scale may be associated with GUT or Planckian physics, and indeed typical realizations of the 
QCD-axion in string theory suggests that $F_{PQ}$ is much closer to the GUT or Planck scale $\Mpl$ \cite{Svrcek:2006yi}, rather than a more intermediate scale, such as $\sim 10^{11}$\,GeV. In some landscape, we can imagine $F_{PQ}$ scanning over different values. For lack of more detailed knowledge, we can imagine that it scans on, say, a uniform distribution in the range $0<F_{PQ}<\Mpl$. If this is the case, then $F_{PQ}$ will be as small as is required, but would not be significantly smaller as that would be even rarer in the landscape. By placing $F_{PQ}$ on a uniform distribution, the probability that it will be small enough to alleviate the instability is roughly
\beq
\mbox{Prob}\,(\mbox{PQ-field alleviates instability})\sim {E^*\over\Mpl}\,,
\label{Prob2}\eeq
where almost all of the phase space pushes $F_{PQ}\sim E^*$, rather than orders of magnitude lower. 
It is important to note that for $E^*\ll\Mpl$, as arises from the measured Standard Model's couplings, the probability in eq.~(\ref{Prob2}) is small but still much greater than the probability in eq.~(\ref{Prob1}). Hence it is much more likely to have an atypically small PQ-scale and no constraint on the initial Higgs field, than an atypically small Higgs field and no constraint on the PQ-scale.
We now examine the cosmological consequences of $F_{PQ}\sim E^*$.

\section{Axion Dark Matter}\label{AxionDM}

The light scalar axion particle is neutral, very stable, and acts as a form of dark matter.
The computation of its abundance is non-trivial and has been studied in many papers, including Refs.~\cite{Preskill:1982cy,Sikivie:2006ni}.
The final result for the axion abundance is essentially controlled by the scale $F_{PQ}$. Its value is normally measured by the quantity $\Omega_{DM}\equiv\rho_{DM}/\rho_{crit}$, where $\rho_{DM}$ is the energy density in axion dark matter and $\rho_{crit}$ is the so-called ``critical density" of the universe defined through the Friedmann equation as $\rho_{crit}=3H^2/(8\pi G)$. 
Tracking the non-trivial temperature dependence of the axion potential and redshifting to late times, leads to the following expression for $\Omega_{DM}$ 
\beq
\Omega_{DM}\approx\chi\,\langle\theta_i^2\rangle\!\left(F_{PQ}\over 10^{12}\,\mbox{GeV}\right)^{\!\!7/6}\!\left(0.7\over h\right)^{\!\!2}
\!\left(T\over 2.725\,\mbox{K}\right)^{\!\!3},
\eeq
where the Hubble parameter is represented as $H=100\,h$\,km/s/Mpc and $T$ is the CMB temperature.
The coefficient $\chi$ is an $\mathcal{O}(0.1-1)$ fudge factor due to uncertainty in the detailed QCD effects that set the axion mass and its temperature dependence. In our numerics we have taken $\chi=0.5$ as a representative value. It is quite possible that the true value may be smaller than this, such as $\chi\approx 0.15$ as taken in Ref.~\cite{Sikivie:2006ni}, but other effects, including contributions from string-axions, etc, can potentially push the true value to be larger \cite{Wantz:2009it}.
Also, $\theta_i$ is the initial $\theta$ angle in the early universe (which later redshifts towards zero, solving the strong CP problem).
Here we take $\langle\theta_i^2\rangle=\pi^2/3$, which comes from allowing $\theta_i$ to be uniformly distributed in the domain $-\pi<\theta_i<\pi$ and then spatial averaging. Another interesting possibility arises if inflation occurs after Peccei-Quinn symmetry breaking, allowing $\theta_i$ to be homogeneous and possibly small, as studied in Refs.~\cite{Pi:1984pv,Tegmark:2005dy}. The latter scenario is subject to various constraints, including bounds on isocurvature fluctuations \cite{Hertzberg:2008wr}, and will not be our focus here.

The quantity $\Omega_{DM}$ is slightly inconvenient for expressing the main results for the following two reasons: (i) in a flat universe (as we are assuming) it is bounded to satisfy $\Omega_{DM}<1$, which obscures the fact that a priori the dark matter abundance could be enormous, and (ii) it is manifestly time dependent (due to $h$ and $T$), which requires some choice of physical time to compare different universes. To avoid these complications, we prefer to compute the dark matter density in units of the baryon density. Fixing the baryon to photon ratio at the measured value, we have 
\beq
\Omega_B\approx \Omega_{B,0} \left(0.7\over h\right)^{\!\!2}\!\left(T\over 2.725\,\mbox{K}\right)^{\!\!3},
\eeq
with $\Omega_{B,0}\approx 0.046$ from observation.
From this we define the (unbounded and time independent) measure of dark matter $\zeta$ as
\bea
\zeta\amp\equiv\amp{\Omega_{DM}\over\Omega_B}={\rho_{DM}\over\rho_B}\,,\\
\amp\approx\amp{\,\chi\,\langle\theta_i^2\rangle\over\Omega_{B,0}}\left(F_{PQ}\over 10^{12}\,\mbox{GeV}\right)^{\!\!7/6} . \label{DMamount}
\eea
Observations show that the dark matter density parameter $\zeta$ is non-zero in our universe, although its particular particle properties (whether axion or WIMP, etc) are still unknown. The observational evidence for dark matter comes from a range of sources, including CMB, lensing, galaxy rotation and clustering, structure formation, baryon-acoustic-oscillations, etc, and is very compelling, e.g., see Refs.~\cite{Bergstrom:2000pn,Primack:2006it,Roos:2010wb,Peebles:2012mz,Carrasco:2012cv,Hertzberg:2012qn}, and its abundance has been measured quite accurately.
Hence our prediction for the value of $\zeta$ (coming from setting $F_{PQ}\sim E^*$ with $E^*$ determined by $m_H$) can be compared to observation; see Fig.~\ref{HiggsAxionPlot}.

\section{Results and Discussion}\label{Discussion}

\subsection{Comparison with Data}

Let us summarize our argument: Holding other parameters fixed, the Higgs mass $m_H$ determines the instability scale $E^*$, which we evaluate at 2-loop order. We have argued on statistical grounds in Section \ref{Meta} why the scale of new physics should not be larger than $E^*$ and in Section \ref{PQDynamics} why the scale of new physics should not be (significantly) smaller than $E^*$, leading to $F_{PQ}\sim E^*$. Since $F_{PQ}$ determines the dark matter abundance $\zeta$ in eq.~(\ref{DMamount}), this establishes a correlation between $m_H$ and $\zeta$. The result was displayed earlier in the paper in Fig.~\ref{HiggsAxionPlot}. The solid-blue curve is for the central value of the top mass $m_t=173.1$\,GeV, and the dashed-blue curves are for $m_t=173.1\pm0.7$\,GeV.
We compare this prediction to the latest LHC and cosmological data. Firstly, we have taken the ATLAS value $m_H=126.0\pm0.4(\mbox{stat})\pm0.4(\mbox{syst})$ \cite{ATLAS}
and the CMS value $m_H=125.3\pm0.4(\mbox{stat})\pm0.5(\mbox{syst})$ \cite{CMS}, and produced our own combined value of $m_H=125.7\pm0.6$\,GeV, 
which is indicated by the red vertical lines.
Secondly, we have taken the WMAP7 data, plus other observations,
for the dark matter abundance $\Omega_{DM}=0.229\pm 0.015$ and the baryon abundance $\Omega_B=0.0458\pm 0.0016$ \cite{WMAP7}
and combined them to obtain $\zeta$, which is indicated by the green horizontal lines. 
The predicted correlation between the Higgs mass $m_H$ and the dark matter abundance $\zeta$ in Fig.~\ref{HiggsAxionPlot} displays good agreement with current data.


\subsection{Precision and Uncertainties}

Improved accuracy in testing this scenario comes in several experimental directions. This includes measuring the Higgs mass $m_H$ to better precision, as well as the top mass $m_t$ and the strong coupling $\alpha_s$ (which we set to the central value $\alpha_s=0.1184$), while the current accuracy in $\zeta$ is quite good. A theoretical uncertainty surrounds the specific choice of $F_{PQ}$ relative to $E^*$. Here we have taken $F_{PQ}\sim E^*$, due to a statistical argument that allowed the scale $F_{PQ}$ to scan, leading to the conclusion that it should be as small as required, but no smaller -- an argument that is similar to the argument for the magnitude of the cosmological constant \cite{Weinberg:1987dv}. One might argue that a factor of a few smaller may be required to properly alleviate the instability \cite{EliasMiro:2012ay}, which would lead to a slight lowering of the blue curves in Fig.~\ref{HiggsAxionPlot}, but a small negative dip is tolerable statistically, which makes $F_{PQ}\sim E^*$ plausible.

Related to this uncertainty is the particular prior distribution for $F_{PQ}$, which we assumed to be uniform. The expectation of a flat distribution is plausible for the cosmological constant $\Lambda$ if one allows both positive and negative values, making $\Lambda\sim 0$ not special. In the case of  $F_{PQ}$, it is necessarily positive, so $F_{PQ}\sim 0$ is arguably a special part of the distribution. This may render the true distribution non-uniform. However, as long as the distribution does not vanish in the $F_{PQ}\to0$ limit faster than $(F_{PQ})^3$, then our arguments go through. In other words, the probability of an atypically small $F_{PQ}$ and no constraint on the initial Higgs field would still be larger than the probability of an atypically small Higgs field and no constraint on $F_{PQ}$. Also, one may question whether the uniform distribution assumed for the initial values of each of the 4 components of the Higgs field is reasonable. Since we have a sufficiently limited understanding of the early universe, including a measure problem for inflation, any such assumptions could be called into question. However, since the meta-stable region occupies such a tiny fraction of the volume of field space, roughly $\sim(10^{-8})^4=10^{-32}$ or so, an alteration in prior probabilities would need to be quite drastic to change the conclusions.

\subsection{Outlook}

An important test of this scenario involves unravelling the nature of dark matter directly.
The QCD-axion is actively being searched for in a range of experiments, including ADMX \cite{Hoskins:2011iv}, with no positive detection so far. But the regime of parameter space in which the axion can be the dark matter will be explored in coming years.
If an axion is discovered, it will be important to unravel its particular properties including its coupling to other fields. An explicit embedding of the discovered version (popular models include KSVZ \cite{Kaxion,SVZaxion} and DFSZ \cite{DFSaxion,Zaxion}) into the Higgs stability analysis would be important. Searches such as ADMX rely upon the axion being all or most of the dark matter, so a related verification would be the associated lack of discovery of WIMPs, or other dark matter candidates, in direct or indirect searches. 
Or at least these forms of dark matter should comprise a relatively small fraction of the total.

The discovery of the Higgs boson at the LHC is a final confirmation of the Standard Model. This leaves the scale at which the theory breaks down unclear. Here we have investigated the possibility that the theory, or at least the Higgs sector, remains intact until the scale at which the Higgs potential runs negative which would lead to a runaway instability at large field values.
By introducing Peccei-Quinn dynamics, we can potentially solve the strong CP problem, remove the unstable region, and obtain roughly the correct amount of dark matter due to a collection of statistical arguments that sets $F_{PQ}\sim E^*$.
This is remarkably minimal, but does still leave questions regarding unification, baryogenesis, inflation, hierarchy problem, etc. 
It is conceivable that unification can still occur at higher energies by the introduction of new degrees of freedom, that the physics of baryogenesis and inflation is associated with such high scale physics \cite{Hertzberg:2011rc,Lyth:1998xn}, 
and that the hierarchy problem has no dynamical explanation. 
Alternatively, the LHC or other experiments may discover new degrees of freedom at much lower energies, which would radically alter this picture.
Currently all such issues remain largely unclear, requiring much more guidance from experiment and observation.

\begin{acknowledgments}
I would like to thank Alan Guth and Frank Wilczek for helpful discussions.
I would also like to acknowledge support from the Center for Theoretical Physics at MIT and the Tufts Institute of Cosmology.
This work is supported by the U.S. Department of Energy under cooperative research agreement Contract Number DE-FG02-05ER41360.
The author declares that there is no conflict of interest regarding the publication of this paper.
\end{acknowledgments}

\section*{Appendix - Standard Model 2-Loop Beta Functions}

In this appendix we list the RG equations for the couplings $\lambda,\,y_t, g',g,g_s$  at energies above the top mass $m_t$ at 2-loop order from Refs.~\cite{Ford,Luo:2002ey}. In each case, we write $d\lambda/dt=\beta_\lambda$, etc, where $t=\ln E/\mu$, and $\mu$ is the starting renormalization scale, taken to be $m_t$. We also performed proper pole matching for couplings defined at the $Z$ mass and running up to the top mass $m_t$, but for brevity do not list those details here.

\begin{widetext}

For the Higgs quartic coupling we have
\bea
\beta_\lambda\amp=\amp
 {1\over(4\pi)^2}  \left[24 \lambda ^2-6 y_t^4+\frac{3}{8} \left(2 g^4+\left(g^2+g'^2\right)^2\right)
 -\left(9 g^2+ 3 g'^2 - 12 y_t^2\right) \lambda \right]\nonumber\\
   \amp\amp+{1\over (4\pi)^4}
  \Bigg{[}\frac{1}{48} \left(915 g^6-289 g^4 g'^2-559 g^2 g'^4-379 g'^6\right)+30
   y_t^6-y_t^4 \left(\frac{8 g'^2}{3}+32 g_s^2+3 \lambda
   \right)\nonumber\\
   \amp\amp+ \lambda  \left(-\frac{73}{8} g^4+\frac{39}{4} g^2 g'^2+\frac{629
   }{24}g'^4+108 g^2  \lambda +36 g'^2 \lambda -312
   \lambda ^2\right)\nonumber\\
   \amp\amp+ y_t^2 \left(-\frac{9}{4} g^4+\frac{21}{2} g^2
   g'^2-\frac{19}{4}g'^4+ \lambda  \left(\frac{45}{2}g^2+\frac{85
   }{6}g'^2+80 g_s^2-144 \lambda \right)\right)\Bigg{]} .
 \eea

For the top quark Yukawa coupling we have
\bea
\beta_{y_t} \amp=\amp
 {y_t\over(4\pi)^2} \left[-\frac{9}{4} g^2-\frac{17
   }{12}g'^2-8 g_s^2+\frac{9}{2}  y_t^2\right]
+{y_t\over(4\pi)^4}
   \Bigg{[}-\frac{23}{4} g^4-\frac{3}{4} g^2
   g'^2+\frac{1187 }{216}g'^4+9 g^2
   g_s^2 \nn\\
 \amp\amp +  \frac{19}{9} g'^2 g_s^2-108
   g_s^4+\left(\frac{225}{16}g^2+\frac{131 }{16}g'^2+36 g_s^2\right)
   y_t^2 + 6 \left(-2 y_t^4-2
   y_t^2 \lambda + \lambda   ^2\right)\Bigg{]} .
   \label{betayt}\eea

For the 3 gauge couplings $g_i=\{g',g,g_s\}$ we have
\bea
\beta_{g_i} \amp = \amp {1\over(4\pi)^2}g_i^3 b_i+{1\over(4\pi)^4}g_i^3\left[\sum_{j=1}^3 B_{ij}g_j^2 - d_i^t y_t^2\right],
\eea
where
\beq
b=\left({41\over6},-{21\over6},-7\right),\quad
B=\left(
\begin{array}{ccc}
199/18 & 9/2 & 44/3 \\
3/2 & 35/6 & 12 \\
11/6 & 9/2 & -26
\end{array}\right),\quad
d^t=\left({17\over6},{3\over2},2\right).
\eeq
By solving the set of 5 coupled differential equations,  we obtain $\lambda$ as a function of energy or $h$.

The wavefunction renormalization of the Higgs field is $G(t)=\exp(-\int_0^t\gamma(t')dt')$, where the anomalous dimension is
\bea
\gamma \amp =\amp  -{1\over(4\pi)^2}  \left[\frac{9 g^2}{4}+\frac{3 g'^2}{4}-3
   y_t^2\right] \nn\\
   \amp\amp - {1\over(4\pi)^4} \left[\frac{271
   }{32}g^4-\frac{9}{16} g^2 g'^2-\frac{431
   }{96}g'^4-\frac{5}{2} \left(\frac{9}{4}g^2+\frac{17
   }{12}g'^2+8 g_s^2\right)
   y_t^2+\frac{27}{4} y_t^4-6
   \lambda ^2\right] .
\eea

\end{widetext}


\begin{thebibliography}{1}

\bibitem{ATLAS} 
  G.~Aad {\it et al.}  [ATLAS Collaboration],
  ``Observation of a new particle in the search for the Standard Model Higgs boson with the ATLAS detector at the LHC,''
  Phys.\ Lett.\ B {\bf 716}, 1 (2012)
  [arXiv:1207.7214 [hep-ex]].
  
  \bibitem{CMS} 
  S.~Chatrchyan {\it et al.}  [CMS Collaboration],
  ``Observation of a new boson at a mass of 125 GeV with the CMS experiment at the LHC,''
  Phys.\ Lett.\ B {\bf 716}, 30 (2012)
  [arXiv:1207.7235 [hep-ex]].

  \bibitem{Aad:2015zhl}
  G.~Aad {\it et al.} [ATLAS and CMS Collaborations],
  ``Combined Measurement of the Higgs Boson Mass in $pp$ Collisions at $\sqrt{s}=7$ and 8 TeV with the ATLAS and CMS Experiments,''
ÊÊPhys.\ Rev.\ Lett.\  {\bf 114} (2015) 191803
Ê
ÊÊ[arXiv:1503.07589 [hep-ex]].       

\bibitem{Sher:1993mf} 
  M.~Sher,
  ``Precise vacuum stability bound in the standard model,''
  Phys.\ Lett.\ B {\bf 317}, 159 (1993)
  [Addendum-ibid.\ B {\bf 331}, 448 (1994)]
  [hep-ph/9307342].

\bibitem{Casas:1994qy} 
  J.~A.~Casas, J.~R.~Espinosa and M.~Quiros,
  ``Improved Higgs mass stability bound in the standard model and implications for supersymmetry,''
  Phys.\ Lett.\ B {\bf 342}, 171 (1995)
  [hep-ph/9409458].
  
\bibitem{PQ1977} R.~D.~Peccei and H.~R.~Quinn, 
``CP conservation in the presence of instantons,"
Phys.~Rev.~Lett. 38, 1440 (1977).

\bibitem{WeinbergAxion} S. Weinberg, 
``A new light boson?,"
Phys.~Rev.~Lett. 40, 223 (1978).

\bibitem{WilczekAxion} F. Wilczek, 
``Problem of strong P and T invariance in the presence of instantons," 
Phys.~Rev.~Lett. 40, 279 (1978).
  
\bibitem{EliasMiro:2012ay} 
  J.~Elias-Miro, J.~R.~Espinosa, G.~F.~Giudice, H.~M.~Lee and A.~Strumia,
  ``Stabilization of the Electroweak Vacuum by a Scalar Threshold Effect,''
  JHEP {\bf 1206}, 031 (2012)
  [arXiv:1203.0237 [hep-ph]].

\bibitem{Lebedev:2012zw} 
  O.~Lebedev,
  ``On Stability of the Electroweak Vacuum and the Higgs Portal,''
  Eur.\ Phys.\ J.\ C {\bf 72}, 2058 (2012)
  [arXiv:1203.0156 [hep-ph]].
  
  \bibitem{Salvio:2015cja} 
  A.~Salvio,
  ``A Simple Motivated Completion of the Standard Model below the Planck Scale: Axions and Right-Handed Neutrinos,''
  Phys.\ Lett.\ B {\bf 743}, 428 (2015)
  [arXiv:1501.03781 [hep-ph]].
  
  \bibitem{Ballesteros:2016euj} 
  G.~Ballesteros, J.~Redondo, A.~Ringwald and C.~Tamarit,
  ``Unifying inflation with the axion, dark matter, baryogenesis and the seesaw mechanism,''
  Phys.\ Rev.\ Lett.\  {\bf 118}, no. 7, 071802 (2017)
  [arXiv:1608.05414 [hep-ph]].
  
\bibitem{WMAP7} 
  E.~Komatsu {\it et al.}  [WMAP Collaboration],
  ``Seven-Year Wilkinson Microwave Anisotropy Probe (WMAP) Observations: Cosmological Interpretation,''
  Astrophys.\ J.\ Suppl.\  {\bf 192}, 18 (2011)
  [arXiv:1001.4538 [astro-ph.CO]].
    
\bibitem{Ford}
 C.~Ford, I.~Jack and D.~R.~T.~Jones,
 ``The Standard Model Effective Potential at Two Loops,''
 Nucl.\ Phys.\  B {\bf 387}, 373 (1992)
 [Erratum-ibid.\  B {\bf 504}, 551 (1997)]
 [arXiv:hep-ph/0111190].

\bibitem{Luo:2002ey} 
  M.~Luo and Y.~Xiao,
  ``Two loop renormalization group equations in the standard model,''
  Phys.\ Rev.\ Lett.\  {\bf 90}, 011601 (2003)
  [hep-ph/0207271].
  
  \bibitem{Buttazzo:2013uya} 
  D.~Buttazzo, G.~Degrassi, P.~P.~Giardino, G.~F.~Giudice, F.~Sala, A.~Salvio and A.~Strumia,
  ``Investigating the near-criticality of the Higgs boson,''
  JHEP {\bf 1312}, 089 (2013)
  [arXiv:1307.3536 [hep-ph]].

\bibitem{Isidori:2001bm} 
  G.~Isidori, G.~Ridolfi and A.~Strumia,
  ``On the metastability of the standard model vacuum,''
  Nucl.\ Phys.\ B {\bf 609}, 387 (2001)
  [hep-ph/0104016].

\bibitem{Degrassi:2012ry}
  G.~Degrassi, S.~Di Vita, J.~Elias-Miro, J.~R.~Espinosa, G.~F.~Giudice, G.~Isidori and A.~Strumia,
  ``Higgs mass and vacuum stability in the Standard Model at NNLO,''
  JHEP {\bf 1208} (2012) 098
  [arXiv:1205.6497 [hep-ph]].
  
\bibitem{Masina:2012tz} 
  I.~Masina,
  ``The Higgs boson and Top quark masses as tests of Electroweak Vacuum Stability,''
  arXiv:1209.0393 [hep-ph].

\bibitem{Svrcek:2006yi} 
  P.~Svrcek and E.~Witten,
  ``Axions In String Theory,''
  JHEP {\bf 0606}, 051 (2006)
  [hep-th/0605206].
   
\bibitem{Preskill:1982cy} 
  J.~Preskill, M.~B.~Wise and F.~Wilczek,
  ``Cosmology of the Invisible Axion,''
  Phys.\ Lett.\ B {\bf 120}, 127 (1983).

\bibitem{Sikivie:2006ni} 
  P.~Sikivie,
  ``Axion Cosmology,''
  Lect.\ Notes Phys.\  {\bf 741}, 19 (2008)
  [astro-ph/0610440].

\bibitem{Wantz:2009it} 
  O.~Wantz and E.~P.~S.~Shellard,
  ``Axion Cosmology Revisited,''
  Phys.\ Rev.\ D {\bf 82}, 123508 (2010)
  [arXiv:0910.1066 [astro-ph.CO]].

\bibitem{Pi:1984pv} 
  S.~-Y.~Pi,
  ``Inflation Without Tears,''
  Phys.\ Rev.\ Lett.\  {\bf 52}, 1725 (1984).

\bibitem{Tegmark:2005dy} 
  M.~Tegmark, A.~Aguirre, M.~Rees and F.~Wilczek,
  ``Dimensionless constants, cosmology and other dark matters,''
  Phys.\ Rev.\ D {\bf 73}, 023505 (2006)
  [astro-ph/0511774].

\bibitem{Hertzberg:2008wr} 
  M.~P.~Hertzberg, M.~Tegmark and F.~Wilczek,
  ``Axion Cosmology and the Energy Scale of Inflation,''
  Phys.\ Rev.\ D {\bf 78}, 083507 (2008)
  [arXiv:0807.1726 [astro-ph]].

\bibitem{Bergstrom:2000pn} 
  L.~Bergstrom,
  ``Nonbaryonic dark matter: Observational evidence and detection methods,''
  Rept.\ Prog.\ Phys.\  {\bf 63}, 793 (2000)
  [hep-ph/0002126].
  
\bibitem{Primack:2006it} 
  J.~R.~Primack,
  ``Precision Cosmology: Successes and Challenges,''
  Nucl.\ Phys.\ Proc.\ Suppl.\  {\bf 173}, 1 (2007)
  [astro-ph/0609541].
 
 \bibitem{Roos:2010wb} 
  M.~Roos,
  ``Dark Matter: The evidence from astronomy, astrophysics and cosmology,''
  arXiv:1001.0316 [astro-ph.CO].
  
\bibitem{Peebles:2012mz} 
  P.~J.~E.~Peebles,
  ``The natural science of cosmology,''
  arXiv:1203.6334 [astro-ph.CO].
  
\bibitem{Carrasco:2012cv} 
  J.~J.~M.~Carrasco, M.~P.~Hertzberg and L.~Senatore,
  ``The Effective Field Theory of Cosmological Large Scale Structures,''
  JHEP {\bf 1209}, 082 (2012)
  [arXiv:1206.2926 [astro-ph.CO]].

\bibitem{Hertzberg:2012qn} 
  M.~P.~Hertzberg,
  ``The Effective Field Theory of Dark Matter and Structure Formation: Semi-Analytical Results,''
  arXiv:1208.0839 [astro-ph.CO].

\bibitem{Weinberg:1987dv} 
  S.~Weinberg,
  ``Anthropic Bound on the Cosmological Constant,''
  Phys.\ Rev.\ Lett.\  {\bf 59}, 2607 (1987).

\bibitem{Hoskins:2011iv} 
  J.~Hoskins, J.~Hwang, C.~Martin, P.~Sikivie, N.~S.~Sullivan, D.~B.~Tanner, M.~Hotz and L.~JRosenberg {\it et al.},
  ``A search for non-virialized axionic dark matter,''
  Phys.\ Rev.\ D {\bf 84}, 121302 (2011)
  [arXiv:1109.4128 [astro-ph.CO]].

\bibitem{Kaxion}
  J.~E.~Kim,
  ``Weak Interaction Singlet and Strong CP Invariance,''
  Phys.\ Rev.\ Lett.\  {\bf 43}, 103 (1979).

\bibitem{SVZaxion} 
  M.~A.~Shifman, A.~I.~Vainshtein and V.~I.~Zakharov,
  ``Can Confinement Ensure Natural CP Invariance of Strong Interactions?,''
  Nucl.\ Phys.\ B {\bf 166}, 493 (1980).

\bibitem{DFSaxion} 
  M.~Dine, W.~Fischler and M.~Srednicki,
  ``A Simple Solution to the Strong CP Problem with a Harmless Axion,''
  Phys.\ Lett.\ B {\bf 104}, 199 (1981).

\bibitem{Zaxion} 
  A.~R.~Zhitnitsky,
  ``On Possible Suppression of the Axion Hadron Interactions. (In Russian),''
  Sov.\ J.\ Nucl.\ Phys.\  {\bf 31}, 260 (1980)
  [Yad.\ Fiz.\  {\bf 31}, 497 (1980)].

\bibitem{Hertzberg:2011rc} 
  M.~P.~Hertzberg,
  ``Can Inflation be Connected to Low Energy Particle Physics?,''
  JCAP {\bf 1208}, 008 (2012)
  [arXiv:1110.5650 [hep-ph]].

\bibitem{Lyth:1998xn} 
  D.~H.~Lyth and A.~Riotto,
  ``Particle physics models of inflation and the cosmological density perturbation,''
  Phys.\ Rept.\  {\bf 314}, 1 (1999)
  [hep-ph/9807278].
        
  

  

  
\end{thebibliography}
\end{document}